\documentclass[12pt]{iopart}
\usepackage{iopams,bbm,amsthm}
\usepackage{epsfig,amssymb,bbm,graphicx}
\usepackage{amsthm,amssymb,epstopdf}

\begin{document}

\title[Continuous-variable Quantum Key Distribution protocols with a discrete modulation]{Continuous-variable Quantum Key Distribution protocols with a discrete modulation}

\author{A Leverrier$^{1,2}$ and P Grangier$^3$}

\address{$^1$ Institut Telecom / Telecom ParisTech, CNRS LTCI, 46, rue
  Barrault, 75634 Paris Cedex 13, France}
  
\address{$^2$ ICFO-Institut de Cienc\`es Fot\`oniques, 08860 Castelldefels (Barcelona), Spain}  
  
\address{$^3$ Laboratoire Charles Fabry, Institut d'Optique, CNRS,
  Universit\'e Paris-Sud, Campus Polytechnique, RD 128, 91127
  Palaiseau Cedex, France}

\ead{anthony.leverrier@icfo.es}

\begin{abstract}
In this paper, we consider continuous-variable quantum key distribution with a discrete modulation, either binary or quaternary. We establish the security of these protocols against the class of collective attacks that induce a linear quantum channel. In particular, all Gaussian attacks are taken into account, as well as linear attacks which add a non-Gaussian noise. We give lower bounds for the secret key rate using extremality properties of Gaussian states. 
\end{abstract}

\maketitle

\section{Introduction}

Quantum Key Distribution (QKD) appears to be the first real-world application of the fastly growing field of quantum information theory \cite{SBC08}. QKD protocols aim at distributing a secret key among distant parties, Alice and Bob, in such a way that an eavesdropper, Eve, cannot learn anything about the key except with an arbitrary small probability $\epsilon$.
Since the proposal of the first QKD protocol in 1984 \cite{BB84}, most schemes have considered encoding information on two-level systems, such as the polarization of a single photon. In such schemes Bob would recover this information with photon counting techniques. 

More recently, homodyne detection has been proposed to replace single photon counters. The main advantages of homodyne detection are its higher quantum efficiency and its greater technological maturity, both being consequences of the fact that homodyne detection is implemented with PIN photodetectors which are commonly used by the telecom industry. This is in sharp contrast with single photon detectors which are almost specific to QKD. In protocols relying on homodyne detection, which we refer to in the following as continuous-variable (CV) protocols by opposition to discrete-variable (DV) protocols, one must encode the information differently than in DV protocols: phase space replaces the traditional Bloch sphere describing the qubits used in BB84 for instance. 
The first CV QKD protocols were exploiting phase space as efficiently as possible by using coherent states with a Gaussian modulation \cite{GG02} which is the modulation maximizing the mutual (classical) information between the input and the output of an Additive White Gaussian Noise (AWGN) channel. Such  a channel models accurately the effect of typical optical fibers on the quadratures of the EM field. Using theorems about the optimality of Gaussian attacks, such schemes were proven secure against collective attacks \cite{NGA06,GC06,LG09b} then later against general attacks \cite{RC09} and were also successfully implemented \cite{LBG07,FDD09}. 

In order to reach significant transmission distances (with more than 3 dB losses),  two different techniques have been  proposed : reverse reconciliation where the key elements are a function of Bob's measurement results \cite{GG02b,GVW03} and post-selection where Alice and Bob discard the data for which Eve has learnt too much information \cite{SRL02}. However, the analysis of post-selected schemes is quite involved their security has been established only against the restricted class of Gaussian attacks \cite{HL07}.
In this paper, we consider protocols with a reverse reconciliation but without post-selection.

In practice,  CV QKD has many advantages on the implementation point of view, but up until now, it seemed restricted to smaller distances than photon-counting QKD, and for instance it was unable to distribute secret keys over more than 50 km. The reason for this comes from the difficulty to correct the errors between Alice and Bob's data induced by the quantum channel. If this is not a problem in principle as one only needs to approach the Shannon capacity of a AWGN channel, it turned out to be quite difficult to realize in practice for a Gaussian modulation \cite{LAB08}. For this reason, using a discrete modulation instead of a continuous Gaussian modulation appeared as a possible solution as it greatly simplifies the error correction step. However, switching from a continuous to a discrete modulation opens new theoretical questions as the proof techniques used in \cite{NGA06,GC06,LG09b} do not apply anymore. One therefore needs to develop a new approach to prove the security of CV QKD protocols with a discrete modulation. This was recently done in \cite{LG09} for a protocol with a quaternary modulation. The new technique allows one to establish the security of the protocol when the quantum channel is linear. In particular, the noise added by the channel can be either Gaussian or non-Gaussian. Moreover, one advantage of the proof is that it only requires to estimate two experimental parameters, namaely the transmission $T$ and excess noise $\xi$ of the quantum channel. Note that several protocols with a discrete modulation had been studied before \cite{LG09}, but their security was only established against Gaussian attacks  \cite{NH03,NH04,LKL04,NH06,HL06,SL10}.  In this paper, we review more in depth the proof of \cite{LG09}, and apply it to analyze a protocol with a binary modulation, considering either homodyne or heterodyne detection schemes. 

The outline of the paper is the following. 
In Section \ref{discrete}, we describe CV QKD protocols with a discrete modulation and we insist particularly on two instances of such protocols for which good error correction schemes are known: the four-state protocol first recently introduced in \cite{LG09} as well as a two-state protocol quite similar to the protocol considered in \cite{ZHR09}. Then, in Section \ref{outline}, we present the general outline of the security proofs of such protocols. We proceed with giving an explicit security proof for each of them, respectively in Sections \ref{two-state} and \ref{four-state}. In Section \ref{rate}, we present the secret key rate of the prootocols and conclude with their expected performances in Section \ref{performance} for realistic, state of the art implementations. In the Appendix, we give some details about the specific reconciliation procedure involved in these protocols.

\section{CV QKD protocols with a discrete modulation}
\label{discrete}

\begin{figure}[!ht]
\centerline{
\begin{tabular}{cc}
\includegraphics[width=0.5\linewidth]{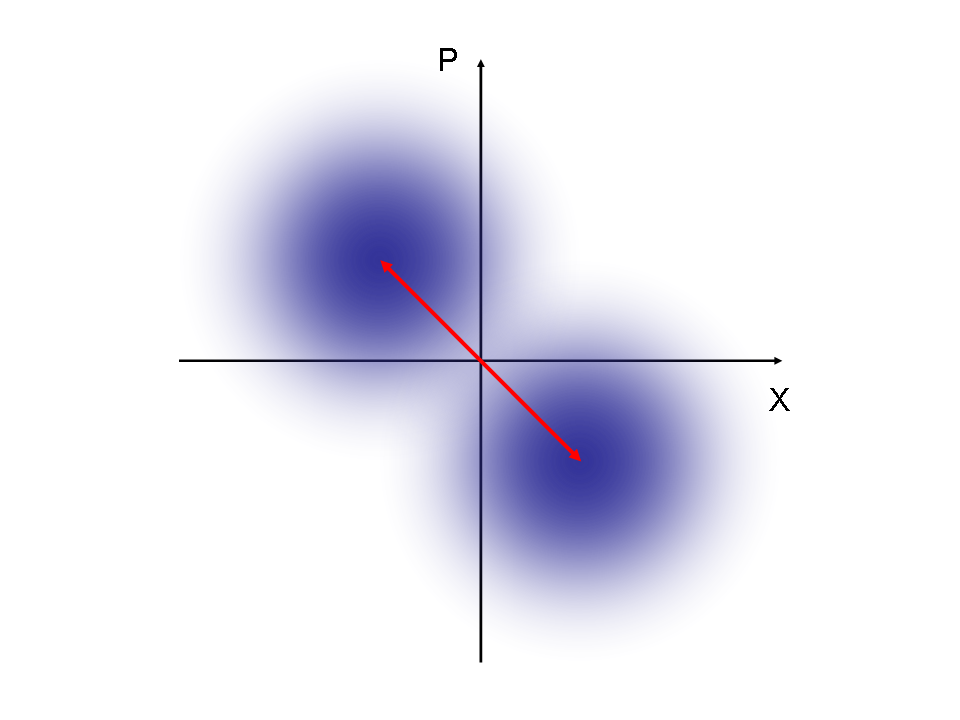} &
\includegraphics[width=0.5\linewidth]{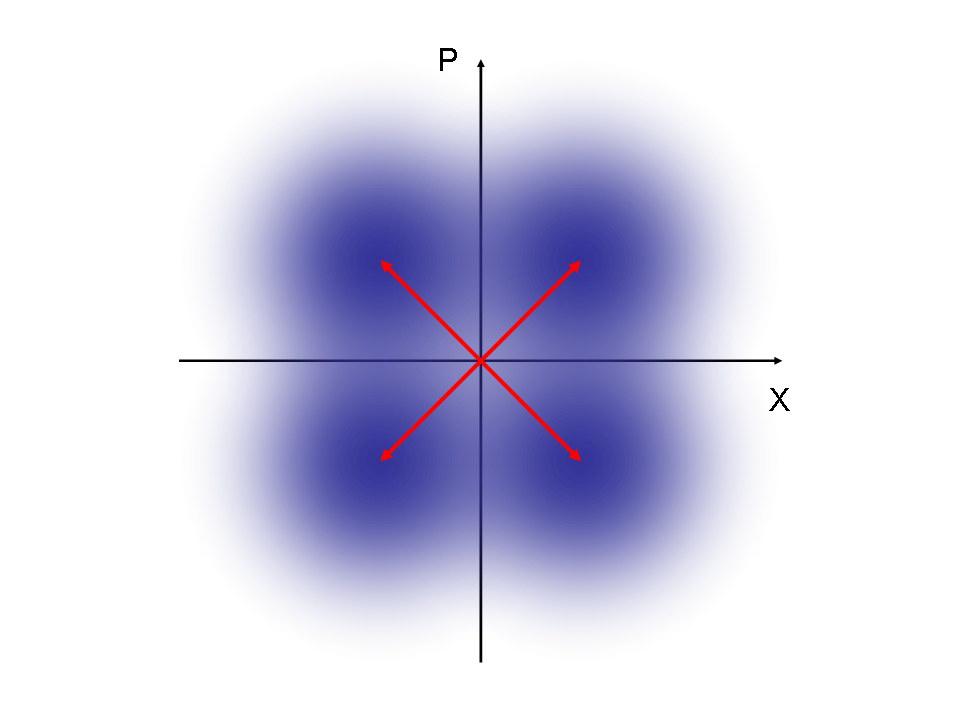} \\
\end{tabular}
}
\caption{\label{modulation} (Color online.) Encoding schemes used for the two-state protocol (left) and the four-state protocol (right).}
\end{figure}

In the following, we consider two CV QKD protocols with a discrete modulation involving respectively two and four coherent states. The four-state protocol was introduced in \cite{LG09}. The two-state protocol is a new protocol which shares similarities with the protocol presented in \cite{ZHR09} but displays a different choice of measurement for Bob. Both modulation schemes are displayed on Figure \ref{modulation}.

In any such protocol, Alice sends $n$ random coherent states drawn from either $\mathcal{S}_2 =  \{|\alpha e^{-i \pi /4}\rangle,|\alpha e^{3i \pi /4}\rangle\}$ or $\mathcal{S}_4 =  \{|\alpha e^{i \pi /4}\rangle,|\alpha e^{3i \pi /4}\rangle,|\alpha e^{5i \pi /4}\rangle,|\alpha e^{7i \pi /4}\rangle\}$ where $\alpha$ is chosen to be a positive real number.  Then, for each state, Bob performs an homodyne measurement on a random quadrature $x$ or $p$. Note that in the case of the two-state protocol, Bob's measurement is not the optimal measurement to distinguish between the state $|\alpha e^{-i \pi /4}\rangle$ and $|-\alpha e^{-i \pi /4}\rangle$. For both protocols, Bob obtains the real random variable $y_i$ for $i \in \{1,\cdots,n\}$.  Alice and Bob use a reverse reconciliation, and the sign $b_i$ of $y_i$ encodes the raw key bit: we note $b_i=1$ if $y_i \geq 0$ and $b_i=0$ if $y_i <0$. Alice must then recover the value of the string ${\bf b} = (b_1, \cdots, b_n)$. To help her, Bob sends some side-information consisting of the quadrature measured, $x$ or $p$, the absolute value of $y_i$ for $i \in \{1,\cdots,n\}$ as well as the syndrome of ${\bf b}$ for a linear error correcting code Alice and Bob agreed on beforehand. Alice then proceeds by decoding her word ${\bf x} = \{x_1, \cdots, x_n\}$ where $x_i$ corresponds to the sign of the quadrature Bob measured for the state she sent. 
An alternative protocol for the four-state protocol consists for Bob to perform an heterodyne (instead of homodyne) measurement \cite{WLB04,WLB06}. In this case, he measures both quadratures, and obtains two measurement results $y_i^x$ and $y_i^p$ for each state sent by Alice. The raw key now consists of both signs of $y_i^x$ and $y_i^p$.

As usual in all QKD schemes,  a supplementary step has to be added for channel estimation: Alice and Bob reveal a fraction of their data and compute the covariance matrix of the state $\rho_{AB}$ that they would share in an entanglement-based version of the protocol \cite{GCW03}. This allows them to compute an upper bound on Eve's information on $y$, the Holevo information $S(y;E)$. If the error correcting code used by Alice and Bob has a rate $R$, the secret key rate against collective attacks (in the limit where the fraction of data revealed for parameter estimation becomes negligible) reads:
\begin{equation}
K = R - S(y;E).
\end{equation} 
$R$ is upper bounded by the mutual information $I(x;y)$ between Alice and Bob and one can therefore introduce the reconciliation efficiency $\beta$ defined as $\beta = R/I(x;y)$. One then finds the more common expression for the secret key rate:
\begin{equation}
K = \beta I(x;y) -S(y;E).
\end{equation}
With this expression, it is clear that being able to perform an efficient reconciliation, with $\beta$ close enough to 1, is crucial. Unfortunately, for a Gaussian modulation, the best reconciliation schemes \cite{BTM06,LAB08} presently known see their efficiency drop under $50 \%$ at low signal to noise ratios (SNR). This is rather dramatic in terms of the range of the protocol as one needs to work at low SNR to distribute secret over long distances. A hand waving argument for this fact is given now. It is known \cite{gro04,NA05} that the quantity $K_{\mathrm{perf}} = I(x;y) - S(y;E)$ tends to a finite limit as the modulation variance $V_A$ of Alice tends to infinity. However, both quantities $I(x;y)$ and $S(y;E)$ diverge to infinity. As a consequence, the penalty $(1-\beta) I(x;y)$ imposed by an imperfect reconciliation also goes to infinity for any value of $\beta$ stricly less than 1. Therefore, one should not work with a large modulation variance as soon as the reconciliation is not perfect, which is never the case. On the contrary, for a realistic reconciliation efficiency (around $80 \%$), the optimal modulation variance is typically quite low (less than 10 photons per pulse on average). Associated with high loss channels, \emph{i.e.} long distance, the SNR is finally very low, well below 1 and reconciliation schemes for Gaussian modulation fail, meaning that no secret key can be exchanged over long distances with such a scheme.

An appealing alternative appears with discrete modulation schemes. The reason for this is that good reconciliation procedures can be found, even at very low SNR, for some modulation schemes. More precisely, in the case of the BI-AWGN channel (where Alice uses a binary modulation $\pm \alpha$ on an AWGN channel), good error correction codes are known at very low SNR \cite{LG09}. Such a binary modulation scheme for error correction can be implemented either with a two-state protocol with coherent states in $\mathcal{S}_2$ or a four-state protocol with coherent states in $\mathcal{S}_4$. In particular, the error correcting codes presented in \cite{LG09} allow one to get a reconciliation efficiency of $80 \%$ for arbitrary low SNR. The main drawback of such protocols is that the methods used to upper bound $S(y;E)$ in the protocol using a Gaussian modulation \cite{NGA06,GC06,LG09b} cannot be directly applied in the case of a discrete modulation scheme. One then needs to come up with a security proof specific to such protocols. We address this question in the next section.

In the remaining of the document, we use the notation $\gamma = \alpha e^{7i \pi/4}  = \alpha e^{-i \pi/4}$ so that the set $\mathcal{S}_2$ of coherent states used in the two-state protocol is $\{|\gamma\rangle, |-\gamma\rangle\}$.

\section{General outline of the security proofs}
\label{outline}

The security of the various protocols we consider here is studied through entanglement-based versions of the protocols. In the \emph{prepare and measure} version of the protocols that are used in practice, Alice randomly draws $n$ binary or quaternary variables, each corresponding to a specific coherent state of $\mathcal{S}_2$ for the two-state protocol or of $\mathcal{S}_4$ in the case of the four-state protocol. Alice then prepares these $n$ coherent states and sends them to Bob through the quantum channel. In the \emph{entanglement-based} version of the protocol, Alice starts with a pure bipartite state $|\Phi_2\rangle$ (or $|\Phi_4\rangle$, depending on the protocol) and performs a projective measurement on the first half of this state. The second half is sent to Bob through the quantum channel. For instance, in the protocol with a Gaussian modulation \cite{GG02}, the initial bipartite state is a two-mode squeezed vacuum, and the projective measurement performed by  Alice is an heterodyne measurement, which projects the second half of the state on a coherent state \cite{GCW03}. The covariance matrix $\Gamma_{\mathrm{TMS}}$ of the two-mode squeezed vacuum reads
 \begin{equation}  
 \Gamma_{\mathrm{TMS}} = 
   \left(
    \begin{array}{cc}
    (1+ 2\alpha^2) \mathbbm{1}_2  & Z_G  \, \sigma_z\\
    Z_G \, \sigma_z &(1 + 2 \alpha^2)  \mathbbm{1}_2\\
  \end{array}
  \right),
\end{equation}
where $\sigma_z = \left( \begin{array}{cc}1&0\\0&-1 \end{array} \right)$ and $Z_G = 2 \sqrt{\alpha^4 + \alpha^2}$. Rewriting $ \Gamma_{\mathrm{TMS}}$ with a direct reference to Alice's modulation variance $V_A$ in the prepare and measure protocol, one has:
\begin{equation}  
 \Gamma_{\mathrm{TMS}} = 
 \left(
    \begin{array}{cc}
    (V_A+1) \mathbbm{1}_2  & Z_G \, \sigma_z\\
    Z_G \, \sigma_z & (V_A+1) \mathbbm{1}_2\\
    \end{array}
  \right).
\end{equation}
As the second half of the state is sent through a quantum channel characterized by its transmission $T$ and excess noise $\xi$, one can write the covariance matrix $ \Gamma_{G}$ of the state $\rho_{AB}$ Alice and Bob share in the CV QKD protocol with a Gaussian modulation:
\begin{equation}  
 \Gamma_{G} = 
  \left(
    \begin{array}{cc}
    (V_A+1) \mathbbm{1}_2  & \sqrt{T} Z_G \,  \sigma_z\\
    \sqrt{T} Z_G \, \sigma_z & (T V_A+1 + T \xi) \mathbbm{1}_2\\
  \end{array}
  \right).
\end{equation}
Then the Holevo information between Eve and Bob's measurement result can be upper bounded by a function of $ \Gamma_{G}$ \cite{GC06}. Note indeed that the argumentation in \cite{GC06} does no rely on the fact that the state considered is indeed Gaussian: only the covariance matrix of the states matters. This is not the case in \cite{NGA06} and \cite{LG09b} where the proof technique explicitely requires the modulation to be Gaussian.

For the protocols of interest in this article, the goal is to apply the same type of proof technique. We therefore want to find a purification $|\Phi_2\rangle$ (resp. $|\Phi_4\rangle$) with a covariance matrix $\Gamma_2$ (resp. $\Gamma_4$) as close as possible as the one of a two-mode squeezed state. This covariance matrix has the following form:
\begin{equation}  
 \Gamma_{2,4} = 
  \left(
    \begin{array}{cc}
    (V_A+1) \mathbbm{1}_2  & Z_{2,4} \, \sigma_z\\
    Z_{2,4} \, \sigma_z & (V_A+1) \mathbbm{1}_2\\
  \end{array}
  \right),
\end{equation}
and the goal is to find a bipartite state $|\Phi_2\rangle$ (resp. $|\Phi_4\rangle$) such that $Z_2$ (resp. $Z_4$) is as close as possible of $Z_G$. Of course, in order to be a legitimate entanglement-based version of the protocol, the bipartite initial state must be such that there exists a projective measurement that Alice can perform that projects the second half of the state onto the desired 2 (or 4) coherent states of the set $\mathcal{S}_2$ (or $\mathcal{S}_4$). Also, it is worth emphasizing that, while the security proof of the Gaussian protocol is based on (virtual) Gaussian entanglement, the present proof  is based on (virtual)  non-Gaussian entanglement, which appears in this context as a convenient theoretical tool.  

Now, the main idea of the discrete modulation protocols we study here is that there exists a regime for the modulation variance, $V_A$, such that 
\begin{equation}
\left\{
\begin{array}{lll}
I_2(x;y) &\approx& I_4(x;y) \approx I_G(x;y),\\
S_2(y;E) &\approx& S_4(y;E) \approx S_G(y;E)\\
\end{array}
\right.
\end{equation}
but with $\beta_2 \approx \beta_4 \gg \beta_G$. Here, the various subscripts $2, 4, G$ refer to the different protocols: the binary modulation, the quaternary modulation and the Gaussian modulation.
The existence of this regime allows one to have $K_2, K_4 >0$ for distances where the secret key rate for a Gaussian modulation $K_G$ is null.

The next two sections are concerned with the study of such states $|\Phi_2\rangle$ and$|\Phi_4\rangle$ and computing the correlation terms $Z_2$ and $Z_4$ of their covariance matrices.

\section{Two-state protocol}
\label{two-state}

We introduce a new QKD protocol involving two coherent state $\{ |\alpha e^{-i \pi/4}\rangle,|-\alpha e^{-i \pi/4}\rangle \}$ where the detection is an homodyne measurement on a randomly chosen quadrature $x$ or $p$. This protocol is quite similar to the protocol studied in \cite{ZHR09} but the difference between the two protocols is that in \cite{ZHR09}, the two coherent states are modulated along one of the quadratures measured by Bob. This is not the case here. The security analysis follows the same lines as for the recently introduced four-state protocol \cite{LG09}.

In the \emph{prepare and measure} version of the protocol, Alice sends the coherent states $\{ |\gamma\rangle $ and $|-\gamma\rangle \}$ with probability $1/2$ to Bob. Hence Bob sees a mixture $\rho_2$ given by:
\begin{eqnarray}
\rho_2 &=& \frac{1}{2} \left(|\gamma\rangle \langle \gamma|+ |-\gamma \rangle \langle -\gamma |\right) \\
&=& \mu_0 |\phi_0\rangle \langle \phi_0| + \mu_1 |\phi_1\rangle \langle \phi_1|,
\end{eqnarray}
where
$\mu_0 = e^{-\alpha^2} \cosh{\alpha^2}$, $\mu_1 = e^{-\alpha^2} \sinh{\alpha^2}$ and
\begin{eqnarray}
|\phi_0\rangle &=& \frac{1}{\sqrt{\cosh{\alpha^2}}} \sum_{n=0}^{\infty} \frac{(-i)^n (\alpha)^{2n}}{\sqrt{(2n)!}} |2n\rangle,\\
|\phi_1\rangle &=& \frac{1}{\sqrt{\sinh{\alpha^2}}} \sum_{n=0}^{\infty} e^{-i \pi/4} \frac{(-i)^n \alpha^{2n+1}}{\sqrt{(2n+1)!}} |2n+1\rangle.
\end{eqnarray}

In order to use the proof technique described in the previous section, we need to consider the \emph{entanglement based} version on the protocol. In this version, Alice starts with a bipartite non-Gaussian pure state $|\Phi_2\rangle$. She performs a projective measurement on one half of the state and sends the other half to Bob through the quantum channel. Depending on the binary result of her measurement, the state sent to Bob is either $|\gamma\rangle $ or $|-\gamma\rangle $ with equal probabilities.
Let us consider the following purification for $\rho$:
\begin{equation}
|\Phi_2\rangle = \sqrt{\mu_0} |\phi_0^*\rangle |\phi_0\rangle + \sqrt{\mu_1} |\phi_1^*\rangle |\phi_1\rangle
\end{equation}
where $|\phi_0^*\rangle$ and $|\phi_1^*\rangle$ are simply defined as:
\begin{eqnarray}
|\phi_0^*\rangle &=& \frac{1}{\sqrt{\cosh{\alpha^2}}} \sum_{n=0}^{\infty} \frac{(i)^n (\alpha)^{2n}}{\sqrt{(2n)!}} |2n\rangle,\\
|\phi_1^*\rangle &=& \frac{1}{\sqrt{\sinh{\alpha^2}}} \sum_{n=0}^{\infty} e^{i \pi/4} \frac{(i)^n \alpha^{2n+1}}{\sqrt{(2n+1)!}} |2n+1\rangle,
\end{eqnarray}
where we recall that $\alpha$ is a positive number.
$|\Phi_2\rangle$ can also be rewritten as:
\begin{equation}
|\Phi_2\rangle = \frac{1}{\sqrt{2}}|\psi_0\rangle |\gamma\rangle +\frac{1}{\sqrt{2}}|\psi_1\rangle |-\gamma\rangle
\end{equation}
with
\begin{equation}
\left\{
\begin{array}{lll}
|\psi_0\rangle &=& \frac{1}{\sqrt{2}} \left(|\phi_0^*\rangle + |\phi_1^*\rangle \right)\\ 
|\psi_1\rangle &=& \frac{1}{\sqrt{2}} \left(|\phi_0^*\rangle - |\phi_1^*\rangle \right)\\
\end{array}
\right.
\end{equation}

\begin{figure}[!ht]
\centerline{
\begin{tabular}{cc}
\includegraphics[width=0.45\linewidth]{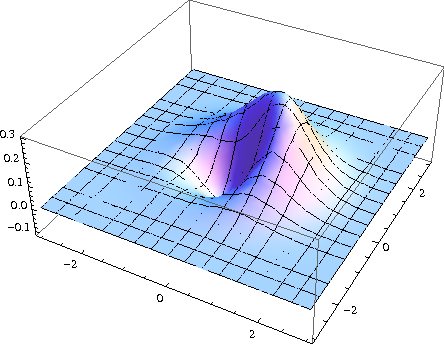} &
\includegraphics[width=0.45\linewidth]{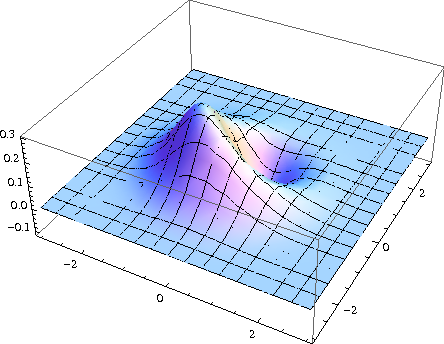} \\
\end{tabular}
}
\caption{\label{cats2} (Color online.) Wigner functions of the states $|\psi_0\rangle$ (left) and $|\psi_1\rangle$ (right) of the two-state protocol for $\alpha^2=0.01$ unit of shot noise.}
\end{figure}

At this point, it is worth noting that in the \emph{entanglement based} version of the protocol, Alice simply applies the projective measurement $\{|\psi_0\rangle \langle \psi_0|,|\psi_1\rangle \langle \psi_1| \}$ to the first half of the state $|\Phi_2\rangle$ and that she therefore projects the second half either on the coherent state $|\gamma\rangle$ or the coherent state $|-\gamma\rangle$ with equal probabilities. The Wigner functions of the orthogonal non-Gaussian states $|\psi_0\rangle$ and $|\psi_1\rangle$ corresponding to Alice's projective measurement are displayed on Figure \ref{cats2}. Note that $|\psi_0\rangle$ is peaked close to the coherent state $|\alpha e^{i \pi/4}\rangle$ while $|\psi_1\rangle$ is peaked close to  $|\alpha e^{5i \pi/4}\rangle$.

We now proceed with evaluating the covariance matrix $\Gamma_2$ of $|\Phi_2\rangle$.
Straightforward algebraic manipulations show that it has the following form:
\begin{equation}  
 \Gamma_2 = 
 \left(
\begin{array}{cc}
    X \mathbbm{1}_2  & Z_2 \, \sigma_z\\
    Z_2 \, \sigma_z & Y \mathbbm{1}_2\\
 \end{array}
\right)
\end{equation}
with
\begin{equation}
\left\{
\begin{array}{lll}
X &= &\langle \Phi_2| 2 a^{\dagger} a +1| \Phi_2\rangle  \\
Y &= &\langle \Phi_2| 2 b^{\dagger} b +1| \Phi_2\rangle  \\
Z_2 &= &\langle \Phi_2| a b + a^{\dagger}b^{\dagger}| \Phi_2\rangle 
\end{array}
\right.
\end{equation}
where $a,a^{\dagger}$ and $b,b^{\dagger}$ are respectively the annihilation and creation operators on Alice and Bob's modes of the state.

In order to compute $X$, one can consider the state $\rho_A = \frac{1}{2} \left(|\gamma^*\rangle \langle \gamma^*|+ |-\gamma^* \rangle \langle -\gamma^*|\right)$ obtained by tracing over the second subsystem of $|\Phi_2\rangle$:
\begin{eqnarray}
X &=& \langle \Phi_2| 2 a^{\dagger} a +1| \Phi_2\rangle\\
&=& \mathrm{tr} (2 a^{\dagger} a +1) \rho_A \\
&=& 1 + \mathrm{tr} (a^{\dagger} a  |\gamma\rangle \langle \gamma|)  + \mathrm{tr} (a^{\dagger} a  |-\gamma\rangle \langle -\gamma|) \\
&=& 1 + 2 \alpha^2
\end{eqnarray}
since $a |\pm \gamma\rangle = \pm \gamma  \; |\pm \gamma\rangle$.
The symmetry of the state $|\Phi_2\rangle$ shows that 
\begin{equation}
Y = \langle \Phi_2| 2 b^{\dagger} b +1| \Phi_2\rangle = X.
\end{equation}
One easily notes that
\begin{equation}
a |\phi_0\rangle = -i \alpha \sqrt{\frac{\mu_1}{\mu_0}} |\phi_1\rangle
\end{equation} 
and
\begin{equation}
a |\phi_1\rangle = i \alpha \sqrt{\frac{\mu_0}{\mu_1}} |\phi_0\rangle.
\end{equation} 
Hence, applying the operator $ab$ on the state $|\Phi_2\rangle$ gives:
\begin{equation}
ab |\Phi_2\rangle =  \alpha^2 \left(\frac{\mu_0}{\sqrt{\mu_1}} |\phi_0^*\rangle |\phi_0\rangle  + \frac{\mu_1}{\sqrt{\mu_0}} |\phi_1^*\rangle |\phi_1\rangle   \right)
\end{equation}
and:
\begin{equation}
\langle \Phi_2 | ab |\Phi_2\rangle =  \alpha^2 \left( \frac{\mu_0^{3/2}}{\mu_1^{1/2}}+ \frac{\mu_1^{3/2}}{\mu_0^{1/2}}  \right) = \alpha^2 \frac{1+e^{-4\alpha^2}}{\sqrt{1-e^{-4\alpha^2}}}
\end{equation}
and finally
\begin{eqnarray}
\langle \Phi_2 | Z_2 | \Phi_2\rangle &=& \langle \Phi_2 | ab + a^{\dagger} b^{\dagger} |\Phi_2 \rangle \\
&=& 2 \mathcal{R}e \langle \Phi_2 | ab |\Phi_2 \rangle \\
&=& 2 \alpha^2 \frac{1+e^{-4\alpha^2}}{\sqrt{1-e^{-4\alpha^2}}}
\end{eqnarray}

The quantity $Z_2$ is displayed on Figure \ref{correlation}. For a variance of modulation less than $0.05$, that is $\alpha \lessapprox 0.15$, $Z_2$ is almost indistinguishable from $Z_G$ thus suggesting that in this regime, $S_2(y;E) \approx S_G(y;E)$. This behaviour will be confirmed in Section \ref{performance}.

\section{Four-state protocol}
\label{four-state}

In this section, we study the protocol recently introduced in \cite{LG09}. More specifically, we introduce a non-Gaussian state $|\Phi_4\rangle$ that can be used in an \emph{entanglement-based} version of the protocol and for which we compute the covariance matrix. 

In the \emph{prepare and measure} version of the protocol, Alice sends the coherent states $\{ |\gamma\rangle, |\gamma^*\rangle , |-\gamma\rangle $ and $|-\gamma^*\rangle \}$ with probability $1/4$ to Bob. Hence Bob sees a mixture $\rho_4$ given by:
\begin{eqnarray}
\rho_4 &=& \frac{1}{4} \left(|\gamma\rangle \langle \gamma|+ |\gamma^* \rangle \langle \gamma^* |+|-\gamma\rangle \langle -\gamma|+ |-\gamma^* \rangle \langle -\gamma^* |\right) \\
&=& \lambda_0 |\phi_0\rangle \langle \phi_0| + \lambda_1 |\phi_1\rangle \langle \phi_1|+\lambda_2 |\phi_2\rangle \langle \phi_2| + \lambda_3 |\phi_3\rangle \langle \phi_3|,
\end{eqnarray}
where
\begin{equation}
\left\{
\begin{array}{lll}
\lambda_{0,2} &= & \frac{1}{2} e^{-\alpha^2}\left( \cosh(\alpha^2) \pm \cos(\alpha^2) \right) \\
\lambda_{1,3} &= & \frac{1}{2} e^{-\alpha^2}\left( \sinh(\alpha^2) \pm \sin(\alpha^2) \right) 
\end{array}
\right.
\end{equation}
and
\begin{equation}
|\phi_k\rangle = \frac{e^{-\alpha^2/2}}{\sqrt{\lambda_k}} \sum_{n=0}^{\infty} (-1)^n \frac{\alpha^{4n+k}}{\sqrt{(4n+k)!}} |4n+k\rangle
\end{equation}
for $k \in \{0,1,2,3\}$.

Applying the annihilation operator $a$ to $|\phi_k\rangle$ gives:
\begin{equation}
a |\phi_k\rangle = \alpha \frac{\sqrt{\lambda_{k-1}}}{\sqrt{\lambda_k}} |\phi_{k-1}\rangle
\end{equation}
for $k \in \{1,2,3\}$ and
\begin{equation}
a |\phi_0\rangle = -\alpha \frac{\sqrt{\lambda_{3}}}{\sqrt{\lambda_0}} |\phi_{3}\rangle.
\end{equation}

\begin{figure}[!ht]
\centerline{
\begin{tabular}{cc}
\includegraphics[width=0.45\linewidth]{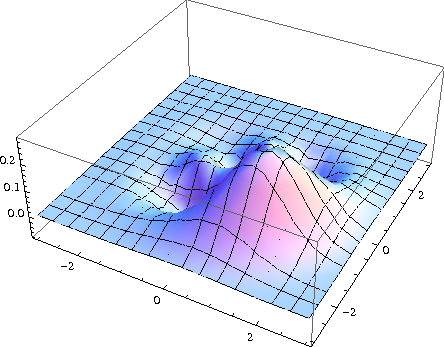} &
\includegraphics[width=0.45\linewidth]{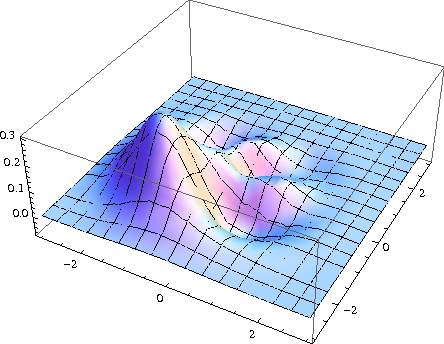} \\
\includegraphics[width=0.45\linewidth]{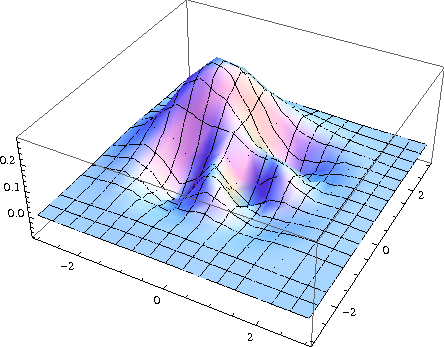} &
\includegraphics[width=0.45\linewidth]{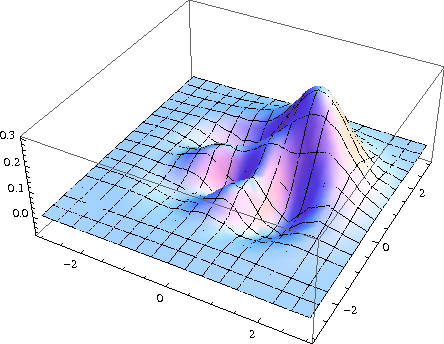}
\end{tabular}
}
\caption{\label{cats} (Color online.) Wigner functions of the states $|\psi_0\rangle, |\psi_1\rangle, |\psi_2\rangle$ and $|\psi_3\rangle$ for $\alpha^2=0.5$ unit of shot noise.}
\end{figure}

Let us now introduce the following purification $|\Phi_4\rangle$ of the state $\rho_4$:
\begin{equation}
|\Phi_4\rangle = \sum_{k=0}^3 \sqrt{\lambda_k} |\phi_k\rangle |\phi_k \rangle.
\end{equation}
This state can also be written as:
\begin{equation}
|\Phi_4\rangle = \frac{1}{2}\left( |\psi_0\rangle |\gamma^*\rangle +|\psi_1\rangle |-\gamma\rangle +|\psi_2\rangle |-\gamma^*\rangle +|\psi_3\rangle |\gamma\rangle  \right)
\end{equation}
where the states 
\begin{equation}
|\psi_k\rangle = \frac{1}{2} \sum_{m=0}^3 e^{i(1+2k)m \pi/4 }|\phi_m\rangle
\end{equation}
are orthogonal non-Gaussian states. These states are displayed on Figure \ref{cats}.

In the \emph{entanglement-based} version of the protocol, Alice needs to perform a projective measurement having  these states as eigenstates, in order to project the second half of $|\Phi_4\rangle$ on one of the four coherent states of $\mathcal{S}_4$, namely $\{|\psi_0\rangle \langle \psi_0|, |\psi_1\rangle \langle \psi_1|, |\psi_2\rangle \langle \psi_2|, |\psi_3\rangle \langle \psi_3|\}$.

Let us compute the covariance matrix $\Gamma_4$ of the bipartite state $|\Phi_4\rangle$. One can show that $\Gamma_4$ has the following form:
\begin{equation}  
 \Gamma_4 = 
  \left(
\begin{array}{cc}
    X \mathbbm{1}_2  & Z_4 \, \sigma_z\\
    Z_4\,  \sigma_z & Y \mathbbm{1}_2\\
 \end{array}
\right)
\end{equation}
where 
\begin{eqnarray}
X=Y&=&\langle \Phi_4 |1+ 2a^{\dagger}a |\Phi_4\rangle = \langle\Phi_4 | 1+2b^{\dagger}b |\Phi_4\rangle\\
&=& \mathrm{tr} ( 1+ 2 a^{\dagger}a \; \rho_4)\\
&=& \mathrm{tr} ( 1+ 2 \sum_{k=0}^3 a^{\dagger}a \; \lambda_k |\phi_k\rangle \langle \phi_k|)\\
&=&  1+ 2 \sum_{k=0}^3 \lambda_k \langle \phi_k| a^{\dagger}a | \phi_k\rangle \\
&=& 1 + 2 \alpha^2 \sum_{k=0}^3 \lambda_k \frac{\lambda_{k-1}}{\lambda_k} \\
&=& 1 + 2 \alpha^2.
\end{eqnarray}

We are now interested in the correlation term of the covariance matrix, that is 
\begin{eqnarray}
\langle \Phi_4 |Z_4|\Phi_4 \rangle &=& \langle \Phi_4 |a b + a^{\dagger} b^{\dagger}|\Phi_4 \rangle\\
&=& 2 \mathcal{R}e \langle \Phi_4 |a b|\Phi_4 \rangle.
\end{eqnarray}
One has:
\begin{eqnarray}
a b |\Phi_4\rangle &=& ab \sum_{k=0}^3 \sqrt{\lambda_k} |\phi_k\rangle |\phi_k\rangle\\
&=& \alpha^2 \sum_{k=0}^3 \frac{\lambda_{k-1}}{\lambda_k} \sqrt{\lambda_k}  |\phi_{k-1}\rangle |\phi_{k-1}\rangle
\end{eqnarray}
where addition should be understood modulo 4. 
Finally, we obtain:
\begin{equation}
\langle \Phi_4 |Z_4|\Phi_4 \rangle = 2 \alpha^2 \sum_{k=0}^3 \frac{\lambda_{k-1}^{3/2}}{\lambda_k^{1/2}}.
\end{equation}

\begin{figure}[ht]
  \centerline{\begin{tabular}{cc}
\includegraphics[width=0.45\linewidth]{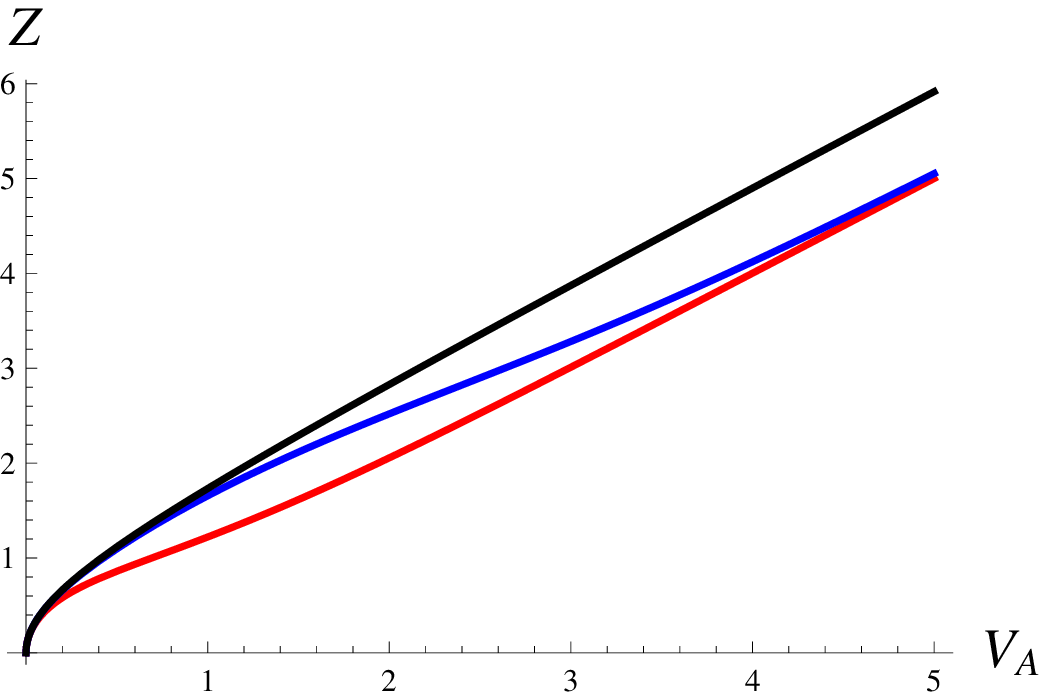} &
\includegraphics[width=0.45\linewidth]{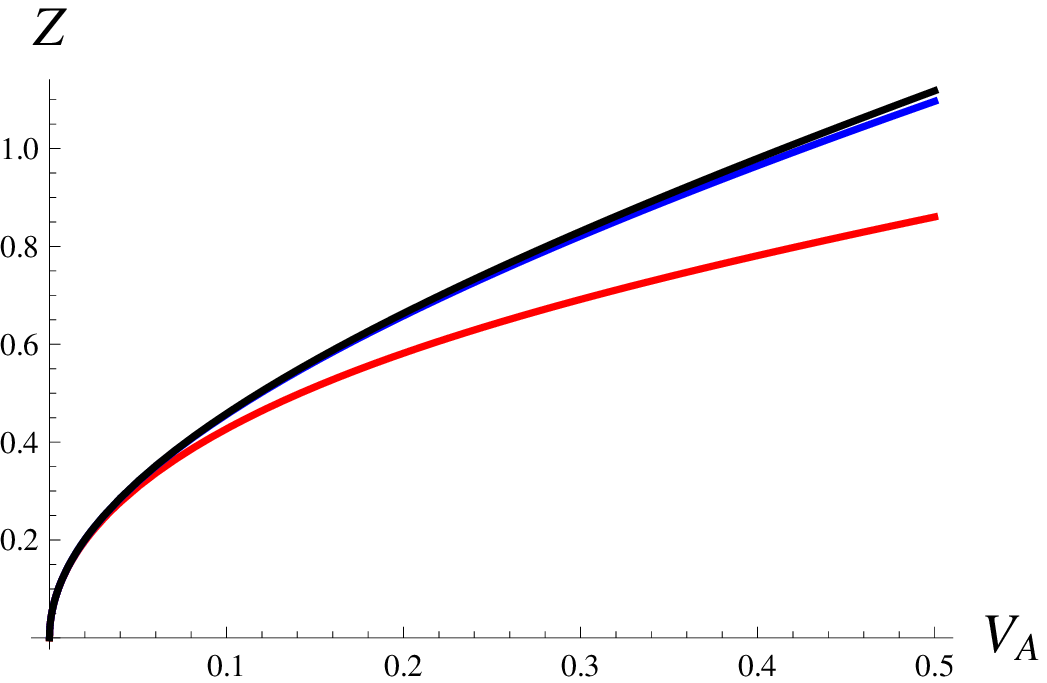} \\
\end{tabular}
}
    \caption{\label{correlation} (Color online.) Comparison of the correlation $Z_2$ for the two-state protocol (lower curve), $Z_4$ for the four-state protocol (middle curve) and for the Gaussian modulation protocol $Z_G$ (upper curve) as a function of the modulation variance $V_A = 2 \alpha^2$. }
\end{figure}

The behaviour of $Z_4$ is plotted on Figure \ref{correlation}. For $V_A \lessapprox 0.5$, that is $\alpha \lessapprox 0.5$, $Z_4$ and $Z_G$ are almost indistinguishable, meaning that in this regime, one has $S_4(y;E) \approx S_G(y;E)$.
We confirm this intuition in the next section.

\section{Secret key rate of the protocols}
\label{rate}

For both protocols, the Holevo information between Eve and Bob's measurement result is upper bounded by the same quantity computed for a Gaussian state $\rho^G_{AB}$ with the same covariance matrix as the state $\rho_{AB}$ shared by Alice and Bob in an entanglement-based version of the protocol. Hence one can give a lower bound for both secret key rates $K_2$ and $K_4$:
\begin{equation}
\left\{
\begin{array}{lll}
K_2 &\geq& \beta I_2(x;y) -  S_2(y;E),\\	
K_4 &\geq& \beta I_4(x;y) -  S_4(y;E).\\
\end{array}
\right.
\end{equation}
The expression for the upper bound on $S_2(y;E)$ (resp. $S_4(y;E)$) is computed from the symplectic eigenvalues $\nu_1, \nu_2$ of $\Gamma_2$ (resp. $\Gamma_4$) and from the eigenvalue $\nu_3$ of the matrix $\Gamma_{2}^{\mathrm{hom}}$ (resp. $\Gamma_{4}^{\mathrm{hom}}$ or $\Gamma_{4}^{\mathrm{het}}$ depending on the nature of the measurement ) corresponding to the covariance matrix of Alice's state given the result $y$ of Bob's homodyne (or heterodyne) measurement \cite{LBG07}.

We computed in the previous sections the covariance matrix of the bipartite state prepared by Alice, that is, \emph{before} the quantum channel. In order to bound Eve's information, we need to know the covariance matrix $\Gamma_{2,4}$ of the state shared by Alice and Bob, that is, \emph{after} the quantum channel.

In this paper, we make the assumption that the quantum channel is \emph{linear} (see \ref{linear} for details). In that case, one can easily use standard techniques from statistics (see Ref. \cite{LGG10}) in order to estimate its transmission $T$ and excess noise $\xi$. 

The covariance matrix $\Gamma_{2,4}$ of the state shared by Alice and Bob is given by:
\begin{equation}  
 \Gamma_{2,4} = 
  \left(
\begin{array}{cc}
    (V_A+1) \mathbbm{1}_2  & \sqrt{T} Z_{2,4} \sigma_z\\
    \sqrt{T} Z_{2,4} \sigma_z & (T V_A+1 + T \xi) \mathbbm{1}_2\\
  \end{array}
\right).
\end{equation}
The reduced covariance matrix given Bob's measurement result depends on the type of measurement performed, either homodyne or heterodyne:
\begin{equation}  
 \Gamma_{2,4}^{\mathrm{hom}} = 
 \left(
\begin{array}{cc}
    V_A+1 - \frac{(Z_{2,4})^2}{T V_A+1 + T \xi}& 0\\
    0 & V_A+1\\
 \end{array}
\right)
\end{equation}
and
\begin{equation}  
 \Gamma_{4}^{\mathrm{het}} = 
\left(
\begin{array}{cc}
    V_A+1 - \frac{(Z_{2,4})^2}{T V_A+2 + T \xi}& 0\\
    0 & V_A+1 - \frac{(Z_{2,4})^2}{T V_A+2 + T \xi}\\
  \end{array}
\right).
\end{equation}

Let us now explain how these covariance matrices can be estimated from experimental data.
Indeed, one should recall that these covariance matrices correspond to a virtual bipartite state, namely the state that Alice and Bob would share in the entanglement-based version of the protocol. Therefore, they cannot be measured directly.
Let us note $x$ and $y$ the respective random variables corresponding to Alice and Bob's classical data in the prepare and measure scenario. One can show that the covariance matrices above can be derived from the observed second moments of the variables $x$ and $y$, that is $\langle x^2\rangle, \langle xy \rangle$ and $\langle y^2 \rangle$.
One has: $V_A = \langle x^2\rangle$, $T = \frac{\langle xy \rangle^2}{\langle x^2\rangle^2}$ and $T V_A+2 + T \xi = \langle y^2 \rangle$.

This shows that the covariance matrices in the entanglement-based scenario are indeed accessible from the experimental data in the actual prepare and measure protocol, assuming that the quantum channel is linear.

\section{Theoretical performances}
\label{performance}

First, it is worth mentioning that the bounds for the Holevo information $S(y;E)$ that we derive from the covariance matrices of $|\Phi_2\rangle$ and $|\Phi_4\rangle$ are not proven to be tight. Indeed, even in the case where the quantum channel between Alice and Bob is perfect, that is, $T=1$ and $\xi=0$, the bounds we compute do not give $S(b;E)=0$ as we would expect, except in the limit of infinitely small modulation variances $\alpha \rightarrow 0$. This is because the states $|\Phi_2\rangle$ and $|\Phi_4\rangle$ are not Gaussian. However, the approximation becomes reasonably good for low modulation variances and one can expect the bounds not to be too loose. 
An intriguing question is whether the value of $S(y;E)$ computed for the Gaussian protocol is an upper bound for the same quantity computed for the discrete-modulation protocols. With the proof we presented, this is not the case (for instance, for a perfect quantum channel, $S_G(y;E)=0$ as expected, whereas the bounds we found for $S_2(y;E)$ and $S_4(y;E)$ are positive). It is quite natural to expect the following relation to hold $S_2(y;E), S_4(y;E) < S_G(y;E)$ since a discrete modulation never maximizes the mutual information between Alice and Bob, and it is doubtful that it presents any advantage for a eavesdropper. However, our security proof cannot bring a definitive answer to this question. 

\begin{figure}[!ht]
  \centerline{
    \includegraphics{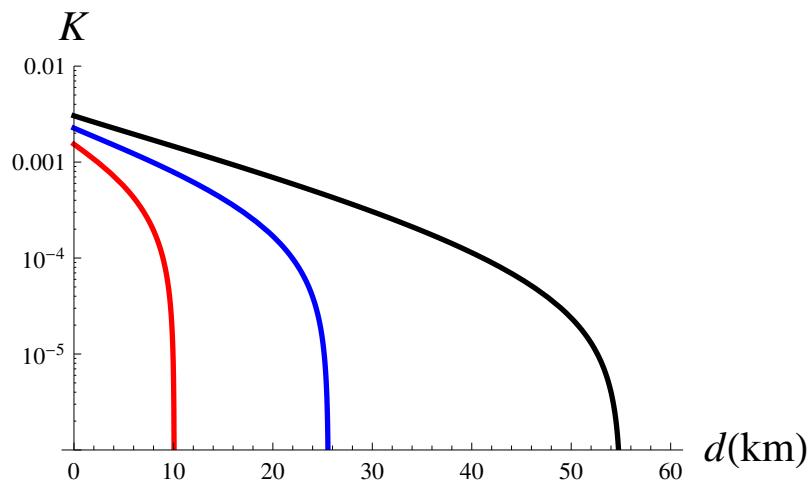}}  
    \caption{\label{perf_imperfect} (Color online.) Secret key rate $K$ of the two-state protocol for a imperfect, realistic reconciliation efficiency of $80\%$ and a quantum efficiency of Bob's detection equal to 0.6. From top to bottom, excess noise is 0.001, 0.0015, 0.002. The respective optimized modulation variances (in number of photons) are 0.015, 0.018 and 0.23. }
\end{figure}

The performances of the two-state protocol are displayed on Figure \ref{perf_imperfect} corresponding to a realistic scheme where the reconciliation efficiency is only $80 \%$ and the quantum efficiency of Bob's detector is equal to $60 \%$ (these values are compatible with state-of-the-art experimental implementations \cite{FDD09}). 
One can see from Figure \ref{perf_imperfect}  that the two-state protocol can only work in a regime where the excess noise is very small: around $1/1000$. We note that this result  is compatible with the results obtained in \cite{ZHR09} where the authors study the security of a slightly different version of the two-state protocol, where they need to assume the perfect knowledge of the probability distribution $p(y|x)$ of Bob's measurements results given Alice's results. 

\begin{figure}[!ht]
  \centerline{
    \includegraphics{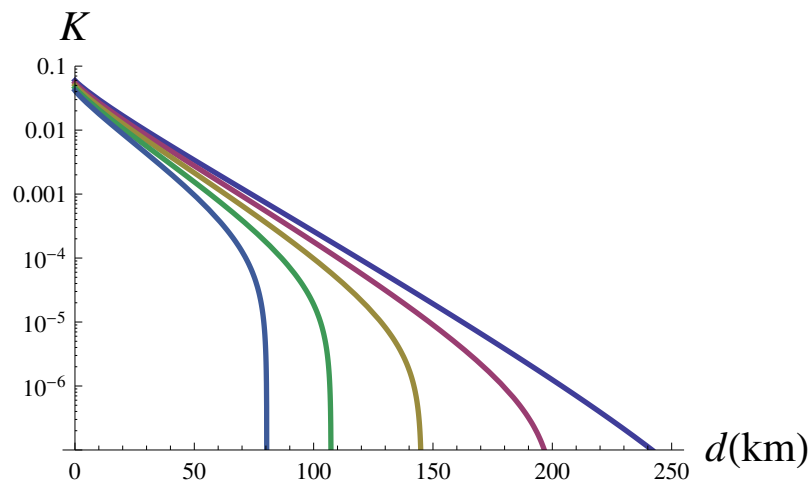}}  
    \caption{\label{perf_4_imperfect} (Color online.) Secret key rate $K$ of the four-state protocol with a homodyne detection for a imperfect, realistic reconciliation efficiency of $80\%$ and a quantum efficiency of Bob's detection equal to 0.6. From top to bottom, excess noise is 0.002, 0.004, 0.006, 0.008 and 0.01. The modulation variance (in number of photons) is 0.125, that is $V_A=0.25$. }
\end{figure}

\begin{figure}[!ht]
  \centerline{
    \includegraphics{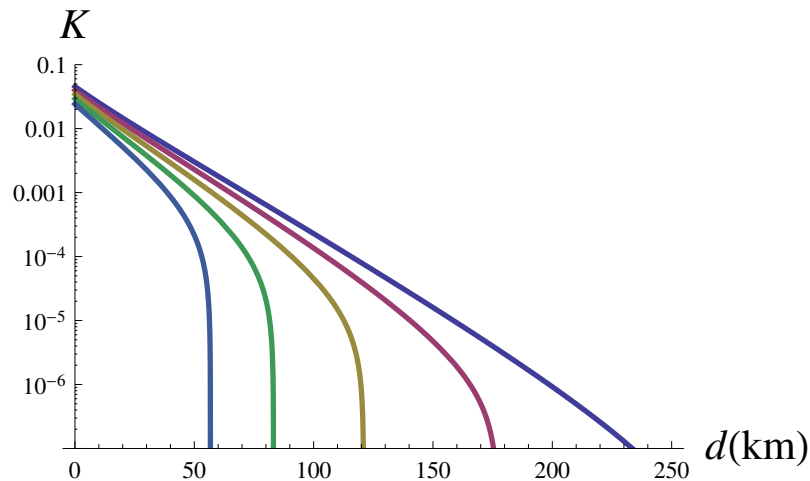}}  
    \caption{\label{perf_4hetero_imperfect} (Color online.) Secret key rate $K$ of the four-state protocol with a heterodyne detection for a imperfect, realistic reconciliation efficiency of $80\%$ and a quantum efficiency of Bob's detection equal to 0.6. From top to bottom, excess noise is 0.002, 0.004, 0.006, 0.008 and 0.01. The modulation variance (in number of photons) is 0.125, that is $V_A=0.25$. }
\end{figure}

The performance of the four-state protocol with a homodyne detection is presented on Figure \ref{perf_4_imperfect} for an realistic reconciliation efficiency of $80\%$ as well as a realistic quantum efficiency of $60\%$ for Bob's detector (which is treated as part of the overall loss between Alice and Bob). The performance of the protocol with a heterodyne detection is displayed on Figure \ref{perf_4hetero_imperfect}. One immediately notices that the four-state protocol performs much better than its two-state counterpart, that is, it allows for a distribution of secret keys over longer distances, and tolerates a much higher (and more reasonable) excess noise. Note that choosing a homodyne or a heterodyne detection does not sensibly affect the performances of the protocol. The better resistance to excess noise of these protocols is extremely important because the results presented so far are a little too optimistic, in the sense that they assume a perfect knowledge of the transmission and excess noise (which is already infinitely less demanding that requiring a perfect knowledge of the quantum channel, which is described by an infinite number of parameters). In practice, however, these parameters can never be perfectly known, and they can only be estimated with a precision depending on the number $N-n$ of data used in the parameter estimation. The main consequence of this imperfect parameter estimation is to increase the effective excess noise, thus decreasing the actual performance of the protocols. The finite size effects for CV QKD protocols are investigated elsewhere \cite{LGG10}.

\section{Conclusion}

We introduced continuous-variable quantum key distribution protocols displaying a discrete modulation. We consider the cases of two and four state modulation. We established the security of these protocols against collective attacks for which the quantum channel is linear. As expected, the four-state protocol clearly outperforms the two-state protocol. The generalization to modulation schemes with a higher number of states may be considered, but it is not so straightforward  as one would lose the main advantage of the protocols presented here, that is, their very efficient reconciliation procedure. However, if one allows for a heterodyne detection instead of a homodyne detection, new continuous modulation schemes can lead to better performances \cite{LG10}.

An important question at that stage is how to avoid the extra hypothesis that the channel should be linear. As shown in Ref. \cite{LG11}, this can be done by using decoy states, in order to embed the non-Gaussian modulation into an overall gaussian modulation. It is then safe to evaluate the values of $T$ and $\xi$ from a gaussian probe beam, and then to use them as described in the present paper.

\section*{Ackowledgements}
This work received financial support from Agence Nationale de la Recherche under projects PROSPIQ (ANR-06-NANO-041-05) and SEQURE (ANR-07-SESU-011-01).

\appendix

\section{Linear quantum channels}
\label{linear}

We shall define a linear quantum channel by the input-output  relations of the quadrature operators in Heisenberg representation : 
\begin{eqnarray}
X_{out} = g_X X_{in} + B_X  \nonumber \\
P_{out} = g_P  P_{in} + B_P
\end{eqnarray}
where the added noises  $B_X$, $B_P$ are uncorrelated with the input quadratures $X_{in}$, $P_{in} $. 
Such relations have been extensively used for instance in the context of Quantum Non-Demolition (QND) measurements of continuous variables \cite{GLP98}, and they are closely related to the linearized approximation commonly used in quantum optics. Gaussian channels (channels that preserve the Gaussianity of the states) are usual examples of linear quantum channels. However, linear quantum channels may also be non-Gaussian, this will be the case for instance if the added noises $B_X$, $B_P$ are non-Gaussian. 

For our purpose, the main advantage of a linear quantum channel is that it will be characterized by transmission coefficients $T_X = g_X^2$, $T_P = g_P^2$, and by the variances of the added noises $B_X$ and $B_P$. These quantities can be determined even if the modulation used by Alice is non-Gaussian, with the same measured values as when the modulation is Gaussian (because these values are intrinsic properties of the channel). The relevant covariance matrix can then be easily determined, and Eve's information can be bounded by using the Gaussian optimality theorem.

\section{Reconciliation at very low SNR}

As we explained, the main reason a Gaussian modulation does no allow for key distribution over very long distances because reconciliation of correlated Gaussian variables is quite complicated at low SNR, and the present techniques are not efficient in this regime.
The main interest of the discrete modulations presented here is that the reconciliation consists in a channel coding problem for the BI-AWGN channel, which turns out to have efficient solutions. In this appendix, we describe in detail this reconciliation procedure.

For both modulation schemes, Bob will ``see'' an effective BI-AWGN channel for either choice of quadrature. Assuming that the quantum channel is known and is indeed Gaussian (which is the case in actual experiments), Alice and Bob can model their classical data respectively as $x=(x_1,\cdots, x_n)$ (with $x_i=\pm \alpha/\sqrt{2}$) and $y=(y_1,\cdots,y_n)$ (here we only consider the data which are used to distill the key, that is, we assume that the parameter estimation was already performed. We also assumed that Bob has informed Alice of his choice of quadrature in the four-state protocol. Therefore, $y_i$ corresponds to Bob's measurement result for the signal $i$ (normalized with the transmission) and $x_i$ corresponds to the corresponding quadrature for Alice's state. The Gaussian channel model reads:
\begin{equation}
y_i = x_i + z_i,
\end{equation}
where $z_i$ is a normal random variable with known variance $\sigma^2$ and $x_i$ is simply an unbiased Bernoulli random variable (that we can assume takes values $+1$ or $-1$ up to a simple renormalisation). 
With these notations, the SNR is given by:
\begin{equation}
\mathrm{SNR} = \frac{1}{\sigma^2},
\end{equation}
and we would like to find reconciliation scheme that perform well, say $\beta = 80 \%$, for very small values of the SNR, for instance $1/100$, or even less. 

\subsection{Good low rate error correcting codes.}
First of all, the reconciliation procedure is necessarily based on good error correcting codes, such as low-density parity-check (LDPC) codes \cite{RSU01}. Despite their great performances, LDPC codes are not universal in the sense that they have not been optimized for every channel. For instance, they perform very well for the BI-AWGN channel when their rate is at least 0.2.

\begin{figure}[!ht]
  \includegraphics[width=.48\linewidth]{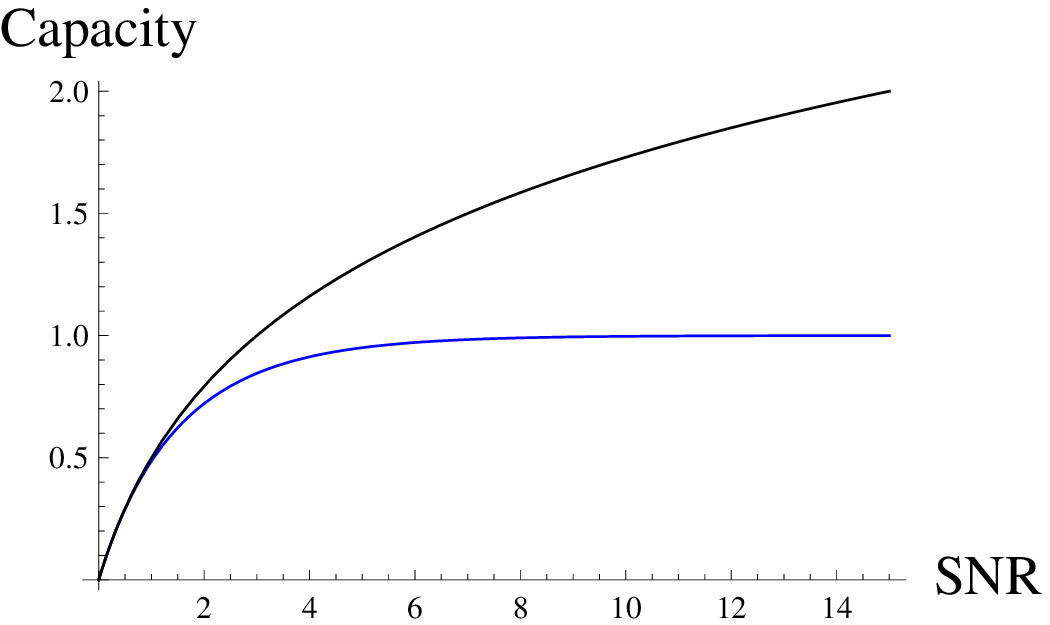}\hfill
  \includegraphics[width=.48\linewidth]{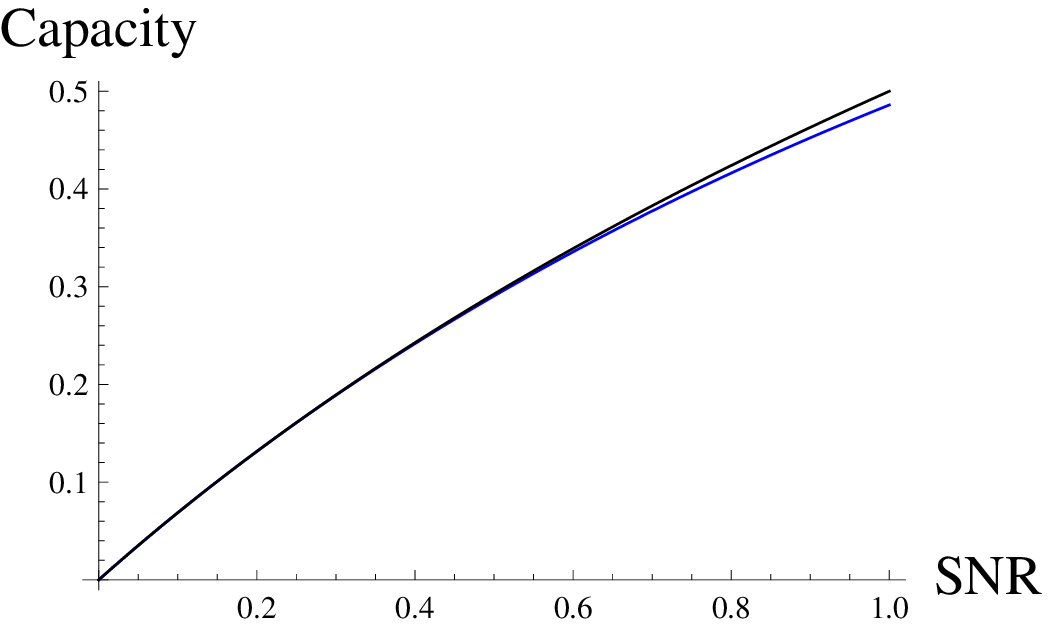}
  \caption{Channel capacities for an AWGN channel with a Gaussian modulation (upper curves) and a binary modulation (lower curves) as a function of the signal-to-noise ratio}
  \label{capacity}
\end{figure}

\begin{figure}[!ht]
 \center{
  \includegraphics[width=.5\linewidth]{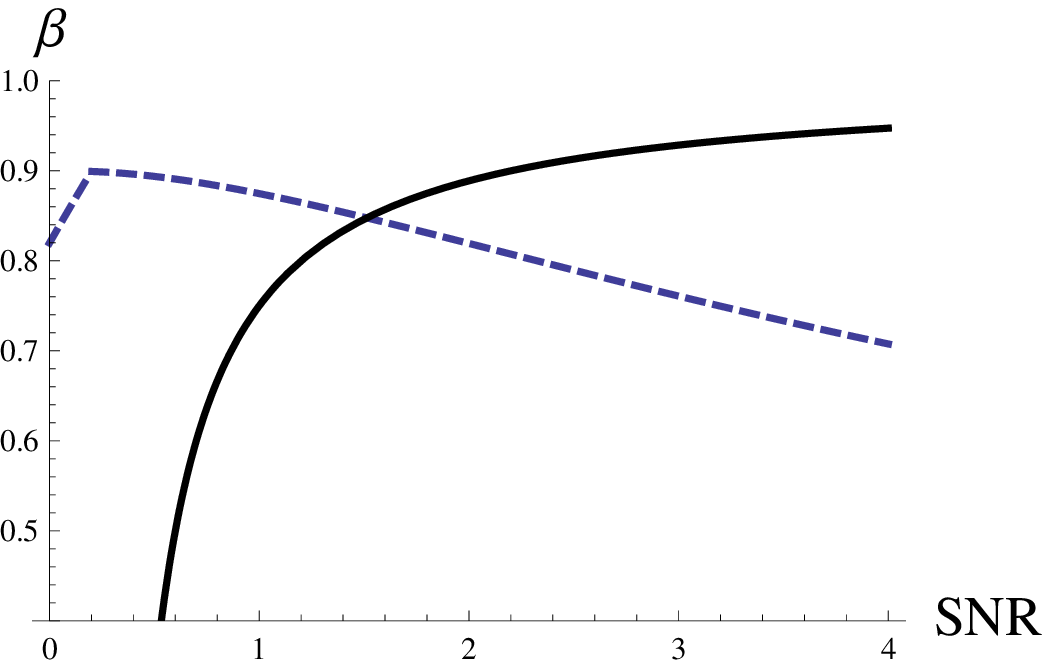}
  \caption{(Color online) Practical reconciliation efficiency for a binary modulation (dashed line) and for a Gaussian modulation (full line)
    \cite{BTM06}.}
  \label{comparison}}
\end{figure}

A special kind of LDPC codes was recently developed to work at reasonably low SNR: the \emph{multi-edge} type LDPC codes \cite{RU02}. Such codes display good performances for rates as low as $1/10$. Even if they help working at low SNR, these codes do not solve our problem completely as we would like codes working at much lower rates. What rate do we need exactly? The rate $R$ is linked to the reconciliation efficiency $\beta$ through
\begin{equation}
\beta = \frac{R}{C_{\mathrm{Gauss}}},
\end{equation}
where
\begin{equation}
C_{\mathrm{Gauss}}= \frac{1}{2} \log_2(1 + s)
\end{equation}
is the capacity of the AWGN channel (which is achieved with a Gaussian modulation) and $s$ is the SNR.
Since in our protocol, we are restricted to a binary modulation, this capacity cannot be reached, and the maximal value of the mutual information between Alice and Bob is given by the capacity of the BI-AWGN channel, $C_{\mathrm{BI-AWGN}}(s)$:
\begin{equation}
C_{\mathrm{BI-AWGN}}(s)=-\int \phi_{s}(x) \log_2(\phi_{s}(x))dx - \frac{1}{2}\log_2(2
\pi e )+\frac{1}{2}\log_2(s)
\end{equation}
where
\begin{equation}
\phi_{s}(x)=\sqrt{\frac{s}{8
   \pi}}\left(e^{-s(x+1)^2/2}+e^{-s(x-1)^2/2}\right).
\end{equation}
Quite interestingly, for small values of the SNR, both quantities $C_{\mathrm{Gauss}}$ and $C_{\mathrm{BI-AWGN}}$ are almost equal as can be seen on Figure \ref{capacity}. However, the two quantities are obviously quite different for large SNR as the Gaussian capacity is unbounded whereas the capacity for a binary modulation is upper bounded by 1: one cannot send more than one bit of information per channel use with a binary modulation. 
With these notations, one can rewrite the reconciliation efficiency as
\begin{equation}
\beta = \beta_{\mathrm{modulation}} \, \frac{R}{C_{\mathrm{BI-AWGN}}},
\end{equation}
where 
\begin{equation}
\beta_{\mathrm{modulation}} = \frac{C_{\mathrm{Gauss}}}{C_{\mathrm{BI-AWGN}}}
\end{equation}
is a factor that rapidly tends to 1 as the signal-to-noise ratio tends to 0, and the second term $R/C_{\mathrm{BI-AWGN}}$ directly reflects the performance of a given code of rate $R$ on the BI-AWGN channel. 
In the limit of low SNR, we can approximate $\beta_{\mathrm{modulation}} \approx 1$, meaning that the code rate that we require is given as a function of the SNR $s$ by
\begin{eqnarray}
R(s) &\approx& \frac{\beta}{2}\log_2(1+ s) \\
& \approx& \frac{\log_2 e }{2} \beta \, s.
\end{eqnarray}
Since we want to fix the value of the reconciliation efficiency (for instance to $80\%$), we see that we need to find error correcting codes with a rate proportional to the signal-to-noise ratio. Hence, we would like to have a process such that if we know a code with rate $R$ and efficiency $\beta$ for a SNR $s$, we can construct a code with rate $R' = R/k$ (for some integer $k \geq 2$) which achieves an efficiency $\beta'$ close to $\beta$ at a SNR $s'=s/k$.
This can be done quite simply with the idea of \emph{repetition code}. 
Let us indeed consider the following scenario: instead of sending a random $x_i = \pm 1$ for each use of the channel, Alice sends $k$ times the same value, that is, $x_{i_1}=x_{i_2} = \cdots =x_{i_k} \equiv X_i$. Hence Bob receives $k$ noisy versions of $X_i$:
\begin{eqnarray}
y_{i_1} &=& x_{i_1} + z_{i_1}\\
y_{i_1} &=& x_{i_1} + z_{i_1}\\
\cdots &=& \cdots \\
y_{i_k} &=& x_{i_k} + z_{i_k},
\end{eqnarray}
where $z_{i_1}, z_{i_2}, \cdots, z_{i_k}$ are $k$ independent and identically distributed random variables: $z_{i_j} \sim \mathcal{N}(0, \sigma^2)$ for $j \in \{1, \cdots, k\}$.
Let us now consider the new random variables defined as:
\begin{equation}
X_i \equiv \frac{1}{k} \sum_{j=1}^k x_{i_j}, \quad
Y_i \equiv \frac{1}{k} \sum_{j=1}^k y_{i_j}, \quad
Z_i \equiv \frac{1}{k} \sum_{j=1}^k z_{i_j}.
\end{equation}
One has
\begin{equation}
Y_i = X_i + Z_i,
\end{equation}
with $X_i = \pm 1$, and $Z_i \sim \mathcal{N}(0, \frac{\sigma^2}{k})$.
The new channel with input $X_i$ and output $Y_i$ is therefore also a BI-AWGN channel but with a signal-to-noise ratio $k$ times higher than for the initial channel. 
Hence, if one knows a code with rate $R$ achieving a reconciliation efficiency $\beta(s)$ for a BI-AWGN channel with SNR $s$, one can use a repetition scheme length $k$ to build a code of rate $R'=R/k$ achieving a reconciliation efficiency $\beta'(s/k)$ for a SNR $s' = s/k$. The new reconciliation efficiency $\beta'(s/k)$ is given by
\begin{equation}
\label{beta_repetition}
\beta'(s/k) = \beta(s) \, \frac{\log_2(1+s)}{k \,\log_2(1+s/k)}.
\end{equation}  
For small values of $s$, this gives $\beta'(s/k) \approx \beta(s)$ as expected. Unfortunately, as we said before, good codes are not known for very small values of $s$, and the best low rate codes presently available are the multi-edge type LDPC codes. In particular, the code of rate $1/10$ described in \cite{RU02} manages to decode reasonably well for a SNR of 0.17. This means that this code is such that $\beta(0.17) \approx 88\%$. 
Using equation \ref{beta_repetition}, one observes that for all $k\geq 1$, $\beta'(0.17/k) \geq 80 \%$. Hence, we can construct codes with arbitrarily low rate that have a reconciliation efficiency greater that $80 \%$. We plot the performance of such codes on Figure \ref{comparison} where we compare it with the reconciliation efficiency achieved with a Gaussian modulation. The difference is striking for low SNR: our concatenation of repetition codes with multi-edge type LDPC codes has a reconciliation efficiency alway greater than $80 \%$ when the SNR tends to zero, whereas the reconciliation efficiency is good (in the sense that in can be used in a CV QKD protocol) only for large enough SNR. 

\subsection{Specificities of the reverse reconciliation.}
Until now, we described a generic method to achieve a good reconciliation efficiency on channels with arbitrarily low SNR. Unfortunately, the approach we described is not directly compatible with QKD. The reason for this is two-fold:
\begin{itemize}
\item first, Alice cannot choose to send $k$ times in a row the same quantum state, as this might give some information to the eavesdropper,
\item but, more importantly, continuous-variable QKD uses \emph{reverse reconciliation}, meaning that Alice needs to guess Bob's measurement result, and not the other way around which would correspond to a direct reconciliation scheme. The problem here is that Bob cannot decide to measure $k$ times in a row the same value. Moreover, it is not completely clear that the channel corresponding to the reverse reconciliation procedure is a BI-AWGN channel as well. We now answer these two points. 
\end{itemize}

\subsection{The reverse reconciliation channel.}
As we said, whereas the direct reconciliation channel is a BI-AWGN channel:
\begin{equation}
\mathrm{input:} \quad x = \pm 1 \; \longrightarrow \; \mathrm{output:} \quad y = x + z \quad \mathrm{with} \quad z \sim \mathcal{N}(0,\sigma^2),
\end{equation}
it is not clear what the reverse reconciliation channel is, simply because its input in real-valued (instead of binary), and that its output is binary instead of being real-valued! In fact, it turns out that this reverse reconciliation channel can be transformed into a BI-AWGN channel, if Bob sends some side-information to Alice. Our goal is to define two variables $u$ for Bob and $v$ for Alice such that the channel mapping $u$ to $v$ is a BI-AWGN channel. 
This can be done through the following procedure. 
First Bob computes two values $u$ and the side-information $t$ from his variable $y$. These two numbers are defined as
\begin{equation}
\left\{
\begin{array}{rcl}
u &=& y/|y|,\\
t &=& |y|
\end{array}
\right.
\end{equation}
Note that for an AWGN channel, the variables $u$ and $t$ are independent: the sign of $y$ is independent from its absolute value since the distribution of $y$ is symmetric. One can also note that $u$ is a unbiased Bernoulli random variable, and therefore corresponds to a legitimate input for a BI-AWGN channel. 
Now, $t$ is considered as a side-information and is sent by Bob to Alice, who can use it to compute a random variable $v$ defined as
\begin{equation}
v = 
\left\{
\begin{array}{c}
t \quad  \mathrm{if} \quad x = 1,\\
- t \quad  \mathrm{if} \quad x = -1.
\end{array}
\right.
\end{equation}
One can check that $u$ and $v$ are related through
\begin{equation}
v = u +w,
\end{equation}
where 
\begin{eqnarray}
w &=& v - u\\
&=& \mathrm{sgn}(x) |y| - \mathrm{sgn}(y)\\
&=& \mathrm{sgn}(y) \, (\mathrm{sgn}(x) y -1) \\
&=& \mathrm{sgn}(y) \, (\mathrm{sgn}(x) (x + z) -1) \\
&=& \mathrm{sgn}(y) \, (1 +\mathrm{sgn}(x)\, z -1) \\
&=& \mathrm{sgn}(xy) \, z 
\end{eqnarray}
which means that $w\sim \mathcal{N}(0, \sigma^2)$ since $\mathrm{Prob}(\mathrm{sgn}(xy)=1) = \mathrm{Prob}(\mathrm{sgn}(xy)=-1) = 1/2$.
Hence, the channel corresponding to the reverse reconciliation scenario, taking $u$ as input and $v$ as output is a BI-AWGN channel.

Let us now show how one can apply the repetition trick to this channel. The main problem now is that one would want $u_{i_1}$ to be equal to $u_{i_2}, \cdots, u_{i_k}$. Obviously, there is only one chance over $2^{k-1}$ for this to happen. The way to overcome this difficulty is in fact quite simple. In the direct reconciliation protocol, Bob would need to guess whether $(x_{i_1}, \cdots, x_{i_k})$ equals $(1, \cdots, 1)$ or $(-1, \cdots, -1)$. In the reverse reconciliation protocol, Bob will inform Alice of the signs of $y_{i_2}, \cdots, y_{i_k}$  relatively to the sign of $y_{i_1}$ (which therefore encode the relevant information), that is, Bob will give Alice the following $(k-1)$ values: $\mathrm{sgn}(y_{i_1}y_{i_2}), \cdots, \mathrm{sgn}(y_{i_1}y_{i_k})$. Hence, in the reverse reconciliation protocol, Alice needs to guess whether $(y_{i_1}, \cdots, y_{i_k})$ equals $(1, y_{i_1}y_{i_2}, \cdots, y_{i_1}y_{i_k})$ or $(-(1, -y_{i_1}y_{i_2}, \cdots, -y_{i_1}y_{i_k})$. Clearly, this problem is completely equivalent to the direct reconciliation case.  In fact, this solution exactly corresponds to Bob informing Alice of the syndrome of his bit string relative to the repetition code of length $k$.

To summarize, the reconciliation procedure starts with Alice and Bob having two correlated vectors of length $k\times m$: $(x_1, \cdots, x_{km})$ and $(y_1, \cdots, y_{km})$. Bob defines the vector ${\bf u} =  (u_1, \cdots, u_{km})$ and sends some side information to Alice, namely the vector ${\bf t} =  (t_1, \cdots, t_{km})$ as well as the $m$ vectors $(1, \mathrm{sgn}(y_{k i+1}y_{k i +2}), \cdots, \mathrm{sgn}(y_{ki + 1}y_{k i +k})$ so that Alice needs to guess the value of the vector ${\bf U} = (\mathrm{sgn}(u_1), \mathrm{sgn}(u_{k+1}), \mathrm{sgn}(u_{2k+1}), \cdots,$ $\mathrm{sgn}(u_{(m-1)k+1}))$, which is a binary vector of length $m$. To do this, Alice and Bob first agree on a particular multi-edge type LPDC code $C$, and Bob sends the syndrome of ${\bf U}$ relative to $C$ to Alice. Alice simply proceeds by decoding $C$ in the coset code defined by the syndrome in question, and recovers ${\bf U}$.  

To conclude, it is easy to adapt the error correction scheme to a reverse reconciliation procedure: it simply involves for Bob to send some well-chosen side-information to Alice through the authenticated classical channel. An important remark is that the role of the side-information is to help Alice to increase the speed and efficiency of  the error correction procedure. The eavesdropper, on the other hand, is supposed to have perfect error correction available and cannot benefit of this side-information as long as it is independent of the key. To see this, let us introduce some additional notations: let $synd$ be the syndrome information that Bob sends to Alice and $|{\bf y}|$ be the vector corresponding to the absolute value of the vector ${\bf y}$. The syndrome $synd$ defines a coset code $C$ for which the word $U$ is a codeword. Let $RK$ be the index of this codeword: this corresponds to the raw key that Alice and Bob will use for privacy amplification.
We want to bound Eve's knowledge on $RK$ given that she has access to the value of $synd$ and $|{\bf y}|$. Using Lemma 1 of Ref. \cite{LAB08} and the fact that $RK$ is independent of both $synd$ and $|{\bf y}|$, one obtains:
\begin{equation}
S(RK; (E, synd, |{\bf y}|)) \leq S(RK, synd, |{\bf y}|; E),
\end{equation}
that is $S(RK; (E, synd, |{\bf y}|)) \leq S({\bf y};E)$ which means that the reconciliation procedure does not give any information to Eve about the raw key $RK$.

The repetition scheme presented above provides a simple method to build a good code of rate $R/k$ out of a code of rate $R$.  This construction is not optimal compared to using a very good error correcting code at the considered signal-to-noise ratio but exhibits some interesting features. First, designing very good codes at low SNR is not easy, and has not been intensively studied so far, mainly because the telecom industry does not operate in this regime: this would not be economical since an important number of physical signals would be required to send one information bit. The problem is very different in QKD, where quantum noise is an advantage rather than a drawback. A second advantage of this repetition scheme lies in its simplicity. As we mentioned earlier, the main bottleneck of CV QKD is the reconciliation : it used to limit both the range and the rate of the protocol. In particular, the rate is limited by the complexity of decoding LDPC codes, which is roughly proportional to the size of the code considered (in fact $O(N  \log N)$). If one uses a repetition scheme of length $k$, then the length of the genuine LDPC code becomes $m=N/k$ allowing a speedup of a factor $k$.  The speed of the reconciliation is not proportional to the number of signals exchanged by Alice and Bob anymore, but to the mutual information they share, which is a major improvement for noisy channels, {\em i.e.}, long distance. Finally, the penalty in terms of reconciliation efficiency imposed by using this scheme instead of a dedicated low rate error correcting code is actually quite small, as soon as one knows a good low rate code. As we saw, a multi-edge type code of rate $1/10$ is sufficient for our purpose.

\section*{References}


\end{document}